\documentclass{ws-procs975x65}
\def\bd{\begin{displaymath}}\def\ed{\end{displaymath}}
\def\be{\begin{equation}}\def\ee{\end{equation}}
\def\bea{\begin{eqnarray}}\def\eea{\end{eqnarray}}
\begin{document}

\title{DOUBLY SPECIAL RELATIVITY AND TRANSLATION INVARIANCE}

\author{S. MIGNEMI$^*$}

\address{Dipartimento di Matematica, Universit\`{a} di Cagliari,\\
09123 Cagliari, Italy\\
$^*$E-mail: smignemi@unica.it}

\begin{abstract}
We propose a new interpretation of doubly special relativity based on the distinction
between the momenta and the translation generators in its phase space realization.
We also argue that the implementation of the theory does not necessarily require a
deformation of the Lorentz symmetry, but only of the translation invariance.
\end{abstract}


\bodymatter
\section{Introduction}
In recent years, the idea that special relativity should be modified for energies
close to the Planck scale $\kappa$, in such a way that $\kappa$ becomes an observer-independent
parameter of the theory, like the speed of light, has been extensively debated \cite{AC,KG,MS}.
This hypothesis is motivated by the consideration that the Planck energy sets a limit
above which quantum gravity effects become important, and its value should therefore not
depend on the specific observer, as would be the case in special relativity.
Of course, this postulate must be implemented in such a way that the principle of
relativity, i.e.\ the equivalence of all inertial observers, still be valid.
The theory based on these assumptions has been named doubly special relativity (DSR)\cite{AC}, 
and implies a deformation of the Poincar\'e invariance and consequently of the dispersion relations
of elementary particles.

Although the most natural implementation of DSR seems to be  the formalism of noncommutative
spacetimes and quantum groups\cite{MR}, an interpretation in terms of classical particle mechanics 
is useful in order to better understand its physical significance. Here we consider the realization 
of DSR in classical phase space and argue that the essence of the formalism relies on the 
distinction between the canonical momenta, interpreted as the physical variables, and the generators 
of translations\cite {Mi1}. This property displays some analogy with the fact that also in curved 
spaces the generators of translations do not coincide with the canonical momenta.

A related observation is that, although DSR is usually associated with the deformation of the 
Lorentz symmetry, through a nonlinear realization of the Lorentz group on momentum space,
its really distinguishing feature is the deformation of the translation symmetry.
In fact, the main phenomenological consequences of DSR \cite{AC} are a deformation
of the addition law of momenta and of the dispersion law of the elementary particles, that
clearly depend only on the nontrivial action of translations.

More specifically, the relation between the momentum spaces of special relativity and DSR can in 
general be obtained by defining the physical momenta $p_\mu$ in terms of auxiliary variables $P_\mu$, 
with $p_\mu=U(P_\mu)$, that satisfy canonical transformation laws under the action of the Poincar\`e 
group \cite{JV}. We show that the physical interpretation of the otherwise obscure auxiliary variables 
can be obtained through their identification with the generators of the 
deformed translations.

To illustrate these considerations, we discuss the Snyder model \cite {Sn} from a DSR point of view.
This model was originally proposed in order to show the possibility of
introducing a noncommutative spacetime without breaking the Lorentz symmetry, and later
interpreted in terms of DSR \cite{KN,Mi1}.

\section{The model}
Let us start by considering the classical action of the Poincar\'e algebra on the phase space of
special relativity.
The Poincar\'e algebra is spanned by the Lorentz generators $J_{\mu\nu}$ and the translation
generators $T_\mu$, obeying Poisson brackets
\bea
\{J_{\mu\nu},J_{\rho\sigma}\}&=&\eta_{\nu\sigma}J_{\mu\rho}-\eta_{\nu\rho}J_{\mu\sigma}+\eta_{\mu\rho}
J_{\nu\sigma}-\eta_{\mu\sigma}J_{\nu\rho},\cr
\{J_{\mu\nu},T_\lambda\}&=&\eta_{\mu\lambda}T_\nu-\eta_{\nu\lambda}T_\mu,\quad\qquad\{T_\mu,T_\nu\}=0.
\eea
Its realization in canonical phase space, with Poisson brackets
\be
\{X_\mu,X_\nu\}=\{P_\mu,P_\nu\}=0,\qquad\{X_\mu,P_\nu\}=\eta_{\mu\nu},
\ee
is obtained through the identification
\be 
J_{\mu\nu}=X_\mu P_\nu-X_\nu P_\mu,\qquad T_\mu=P_\mu,\ee
which yields the infinitesimal transformation laws for the phase space coordinates $X_\mu$ and $P_\mu$,
\be
\{J_{{\mu\nu}},X_\lambda\}=\eta_{\mu\lambda}X_\nu-\eta_{\nu\lambda}X_\mu,\qquad
\{J_{\mu\nu},P_\lambda\}=\eta_{\mu\lambda}P_\nu-\eta_{\nu\lambda}P_\mu,\ee
\be
\qquad\{T_\mu,X_\nu\}=\eta_{\mu\nu},\qquad\{T_\mu,P_\nu\}=0.
\ee

As discussed previously, to derive the transformation laws of DSR under the action of the 
Poincar\'e group, one introduces the physical momenta $p_\mu$ as a nonlinear function of the 
variables $P_\mu$ that obey the standard transformation laws of special relativity.
In the case of the Snyder model, this relation reads
\be\label{sny}
p_\mu=U(P_\mu)={P_\mu\over\sqrt{1+\Omega P^2}},
\ee
where $\Omega=1/\kappa^2$ is the Planck area.

In a DSR interpretation, the Snyder model can then be characterized by the explicitly Lorentz
invariant deformed dispersion relation $p^2/(1-\Omega p^2)=m^2$, obtained by substituting 
(\ref{sny}) into the standard deformation relation $P^2=m^2$. This can also be written 
$p^2=m^2/(1+\Omega\,m^2)$.
In this form, the dispersion relation looks like a redefinition of the mass (notice that
the dispersion relation for massless particles maintains its classical form); nevertheless,
as we shall see, some nontrivial consequences follow.
From the structure of (\ref{sny}) it is instead evident that the action of the Lorentz 
group on the momentum variables is not affected.

A full description of physics requires the definition of a spacetime structure compatible with 
the deformation of the Poincar\'e invariance. It is therefore natural to introduce position
variables $x_\mu$ that transform covariantly with respect to the momenta. These can be
defined as \cite{Mi2,Mi1}
\be
x_\mu=\sqrt{1+\Omega P^2}\,X_\mu.
\ee
With this definition, the Poisson brackets between the new phase space coordinates are 
no longer canonical, and the position space becomes noncommutative, realizing the 
proposal of Snyder\cite{Sn},
\be
\{x_\mu,x_\nu\}=-\Omega(x_\mu p_\nu-x_\nu p_\mu),\quad\{p_\mu,p_\nu\}=0,
\quad\{x_\mu,p_\nu\}=\eta_{\mu\nu}-\Omega\,p_\mu p_\nu.
\ee

In terms of the physical coordinates $x_\mu$ and $p_\mu$, the generators of the 
Poincar\`e group read
\be
J_{\mu\nu}=x_\mu p_\nu-x_\nu p_\mu,\qquad T_\mu=P_\mu={p_\mu\over\sqrt{1-\Omega p^2}},
\ee
where we have identified the translation generators with the auxiliary variables $P_\mu$.
The transformation laws of $x_\mu$ and $p_\mu$ under the Lorentz subalgebra maintain the 
canonical form, while under translations become
\be\label{tl}
\{T_\mu,x_\nu\}={\eta_{\mu\nu}\over\sqrt{1-\Omega p^2}},\qquad\{T_\mu,p_\nu\}=0.
\ee
Therefore, the effect of the translations on the position coordinates becomes momentum
dependent and increases for near Planck-mass particles.

It can also be shown that the sum rule of momenta is modified.
Moreover, it is possible to define a spacetime metric which is invariant under the 
transformations (\ref{tl}). This is given by $ds^2=(1-\Omega p^2)dx^2$ and, as usual in DSR, 
depends explicitly on the momentum. More details on these topics can be found in 
Ref.~\refcite{Mi1}.

\end{document}